\documentclass{optica-article}
\journal{opticajournal} 
\articletype{Research Article}
\usepackage{xcolor}
\usepackage{tikz} 
\usepackage{pgfplots,pstricks}
\pgfplotsset{compat=1.18}
\usepackage{lineno}
\usepackage{array}
\usepackage{amsmath}
\usepackage{caption}
\usepackage{gensymb}
\usepackage{subcaption}
\usepackage{graphicx}
\definecolor{option1}{HTML}{DF6D14}
\definecolor{mycolor4}{rgb}{0.54117,0.78823,0.14902}%
\graphicspath{ {./Figures/} }
\begin{document}
\title{Photocurrent detected 2D spectroscopy via pulse shaper: insights and strategies for optimally untangling the nonlinear response}
\author{Edoardo Amarotti,\authormark{1} Luca Bolzonello,\authormark{2} Sun-Ho Lee,\authormark{3} Donatas Zigmantas,\authormark{1} Nam-Gyu Park,\authormark{3,4} Niek van Hulst,\authormark{2,5} and Tönu Pullerits\authormark{1,*}}
\address{%
\authormark{1}Department of Chemical Physics and NanoLund, Lund University, P.O. Box 124, 22100 Lund, Sweden\\
\authormark{2}ICFO - Institut de Ciencies Fotoniques, The Barcelona Institute of Science and Technology, Castelldefels, Barcelona, 08860 Spain\\
\authormark{3}School of Chemical Engineering, Sungkyunkwan University, Suwon 16419, Republic of Korea\\
\authormark{4}SKKU Institute of Energy Science and Technology (SIEST), Sungkyunkwan University, Suwon 16419, Republic of Korea\\
\authormark{5}ICREA—Institució Catalana de Recerca i Estudis Avançats, Barcelona 08010, Spain}
\email{\authormark{*}tonu.pullerits@chemphys.lu.se} 
%
%
\begin{abstract*} 
Action-detected two-dimensional electronic spectroscopy (A-2DES) provides valuable insights into ultrafast dynamics within functional materials and devices by measuring incoherent signals like photocurrent. This work details the implementation and optimization of a pulse-shaper-based A-2DES setup, focusing on methodological strategies crucial for acquiring high-fidelity data. We present a comprehensive analysis of phase modulation routines, elucidating the critical interplay between pattern parameters (N, n\textsubscript{i}), pattern repetitions (N\textsubscript{rep}), laser repetition rate, and acousto-optic pulse shaper constraints (e.g., streaming rate, RF generator nonlinearities). Utilizing a perovskite solar cell as a model system, we systematically identify and characterize significant inaccuracies inherent to A-2DES measurements. These include distortions originating from Fourier transform processing of improperly trimmed time-domain data (phase leakage), signal accumulation effects due to insufficient sample response discharge between pulse sequences at high repetition rates, and shortcomings induced by pulse shaper operation at elevated streaming powers. Crucially, we demonstrate robust data post-processing strategies, including precise data point selection for Fourier analysis and phase correction routine, to effectively mitigate these imperfections and retrieve accurate 2D spectra. This rigorous methodological investigation and anomalous features characterization provides essential guidelines for optimizing pulse-shaper-based A-2DES experiments, ensuring data integrity and enabling reliable extraction of complex photophysical information in complex systems.
\end{abstract*}
%
\section{Introduction}
Ultrafast spectroscopy is a powerful tool for unravelling the intricate physical properties of complex systems and advanced materials. By enabling the direct observation of ultrafast dynamics on femtosecond to picosecond timescales, these techniques provide deep insights into a diverse range of fundamental phenomena, including molecular vibrations, electronic transitions, coupling mechanisms, and energy transfer pathways \cite{Mukamel2000,Jonas2003}. Among the various ultrafast spectroscopic techniques, multidimensional spectroscopy has emerged as a particularly valuable approach due to its unparalleled ability to achieve high temporal and spectral resolution simultaneously \cite{Hamm2011,Zigmantas2006}. This capability facilitates a more comprehensive understanding of the underlying interactions governing energy flow and charge carrier dynamics in complex systems. Multidimensional spectroscopy can be broadly categorized based on whether the detected signal is collected as a coherent or incoherent response. Traditional methods, such as transient absorption (TA) and two-dimensional electronic spectroscopy (2DES), typically rely on the measurement of coherent signals generated through nonlinear interactions with multiple light pulses in a specific phase-matching direction \cite{Mukamel1995}. These techniques have provided a wealth of information about ultrafast processes and are widely applied to study excited-state dynamics of interest. However, these methods can face inherent limitations when applied to certain systems. As an example, signal coming from highly absorptive samples is often too weak to measure accurately with coherent techniques as TA spectroscopy; another common issue is dealing with complex systems: mixtures or composite materials may produce signals from different components, complicating the extraction of precise information; an additional problem might arise from the sample’s physical state: solid samples, especially those that are opaque or thick, can strongly scatter light, while highly reflective surfaces or materials with low transparency may distort the incoming light, preventing accurate measurement.

In this context, a powerful technique that can deal with these challenges is the action-detected two-dimensional spectroscopy (A-2DES). A-2DES is an advanced ultrafast spectroscopic technique that, unlike conventional TA or 2DES, detects an incoherent signal linked to a specific observable \textemdash \ such as fluorescence \cite{Malý2020,Mueller2018,Tekavec2007,Tiwari2018,Wagner2005,Karki2019} or photocurrent \cite{Bakulin2016,Bolzonello2021,Karki2014,Nardin2013} \textemdash \ following photoexcitation by four sequentially delayed optical pulses. By controlling the phases of the four pulses through a phase cycling \cite{Tan2008,Tian2003} or phase modulation \cite{Tekavec2007,Tekavec2006} scheme, it is possible to retrieve high-resolution frequency information along both the excitation and detection axes, allowing for the extraction of the nonlinear response via Fourier transform (FT) analysis of the signal related to the fourth order excited state population \cite{Nardin2013}. The implementation of A-2DES requires precise control over experimental parameters to ensure high-quality and reproducible results. Factors such as pulse timing, phase stabilization, and detection sensitivity must be meticulously optimized to extract meaningful spectral features and accurately interpret underlying physical mechanisms \cite{Fresch2023}.

In this study, we present a comprehensive overview of the pulse shaper-based experimental setup and methodological advancements concerning the data post-processing developed for A-2DES measurements. We highlight critical yet often overlooked factors that are crucial for obtaining reliable spectroscopic data spanning from optimization of the pulse shaper working conditions to precise data trimming for optimal FT during the post-processing data handling, avoiding imprecisions that can introduce undesired distorted features in the 2D map. Our results, recorded on perovskite solar cell devices at room temperature, demonstrate the potential of A-2DES for advanced material characterization.
\section{Materials and methods}
\subsection{Photocurrent-detected 2D electronic spectroscopy setup}
\begin{figure}[htbp]
\centering
\includegraphics{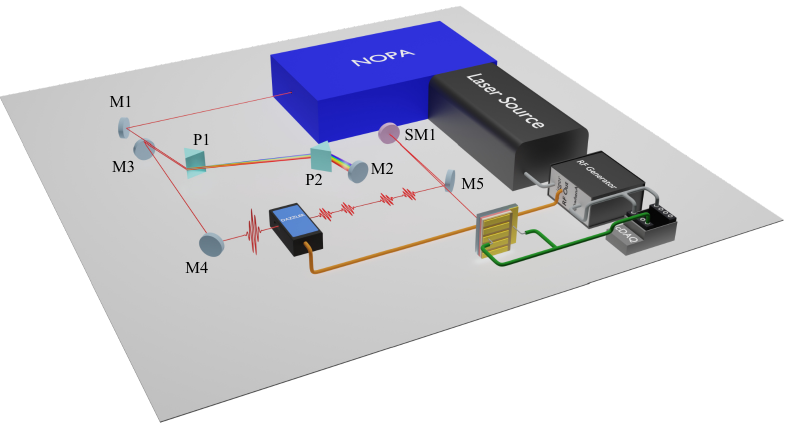}
\caption{Setup schematic: M = flat mirror; P = prism; SM = spherical mirror. After NOPA, the pulse is pre-chirped through a prism compressor and sent into the Dazzler pulse shaper. The 4 collinear replicas generated are then focused using a spherical mirror directly into the sample. The photocurrent signal is then collected through a cDAQ system.}
\label{Fig1}
\end{figure}

The experimental setup developed is sketched in Fig. \ref{Fig1}, and it is inspired by the design of Bolzonello et al. \cite{Bolzonello2021}. A Yb:KGW amplified laser system (Pharos, Light Conversion Ltd.) generates 170 fs light pulses at 1030 nm, with a repetition rate ranging from 1 up to 200 kHz. A custom-built non-collinear optical parametric amplifier (NOPA) spectrally broaden the incoming laser pulses into the visible and near-infrared spectral range (470 to 900 nm) with an energy per each excitation pulse up to 10 nJ. Subsequently, the pulses are pre-chirped through a prism-based compression system (LaKL21, Layertec GmbH) and directly sent into an acousto-optic pulse shaper (Dazzler, Fastlite) \cite{Tournois1997,Verluise2000a,Verluise2000b}. Assuming a laser bandwidth of 100 nm, the Dazzler can generate the shaped pulses with a pulse duration down to 15-20 fs, enabling a scanning range for the population time up to 2 ps. Taking advantage of the streaming option of the Dazzler, each incoming pulse is diffracted into a set of 4 fully collinear pulses when interacting with the acoustic waveform streamed into its birefringent crystal. By controlling the shape of the acoustic waveform, a precise control over both the time-delay between each pulse and their individual phase is achieved. The set of 4 collinear phase-modulated pulses are then focused onto the sample by a spherical mirror down to a beam waist of 100 \textmu m. The detection system is composed by a National Instruments CompactDAQ USB chassis (NI cDAQ‑9174) containing the NI-9215 analog input module, which collects the photocurrent response, and a National Instruments digital I/O module (NI-9402) to trigger the signal acquisition. All the experimental setup is controlled by a home-built interface based on 
LabVIEW software.
\subsection{Phase modulation routine}
Within a theoretical framework, phase modulation provides the flexibility to select any combination of four distinct frequencies to modulate the phase of the four collinear pulses generated by the pulse shaper. However, the interplay between the pulse shaper and the FT imposes inherent constraints on this selection. First, due to the limitations of the FIFO (First In, First Out) memory in our Dazzler system, a maximum of 41 different waveforms can be uploaded per streaming session. Additionally, to ensure optimal FT performance during post-processing, the phase evolution must adhere to a smooth, gradual transition over the course of signal acquisition. This requirement is dictated by the sinusoidal modulation associated with the specific frequency assigned to each pulse, making it essential to avoid abrupt phase shifts and discontinuities. To satisfy these criteria, the experimental protocol utilizes a pattern of N waveforms, each designed to diffract the incoming laser pulse into four replicas with precisely controlled time delays and phases, key parameters that dictate the experimental outcome. By systematically varying the time delays and recording the corresponding signal response, a time-domain 2D map can be constructed. Performing a two-dimensional Fourier transform (2D-FT) then resolves the excitation and detection axes in the frequency-domain, providing a powerful tool to resolve overlapping spectral features, identify couplings, and track ultrafast dynamics. Control over the phase of each diffracted pulse enables the retrieval of the fourth-order signal by applying a FT to the action-detected response in contrast with conventional 2D spectroscopy, where the frequency-resolved third-order signal is coherently detected directly along the designated phase-matching direction. More in detail, each of the four diffracted pulses in the set is modulated by a mathematically independent frequency, determined by selecting n\textsubscript{i}, a unique integer divisor of N (the pattern size) for each pulse. This choice ensures that for each laser shot, the phase advances by:
\begin{equation}
\Delta\Phi_\mathrm{i} = \mathrm{n}_\mathrm{i}\cdot\frac{2\pi}{\mathrm{N}}
\label{Eq1}
\end{equation}

where the index i = 1,2,3,4 refers to a different pulse in the diffracted set.

\begin{figure}[htbp]
\centering
\includegraphics{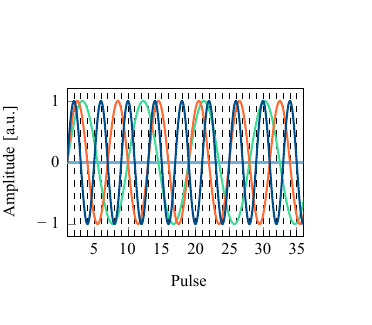}
\includegraphics{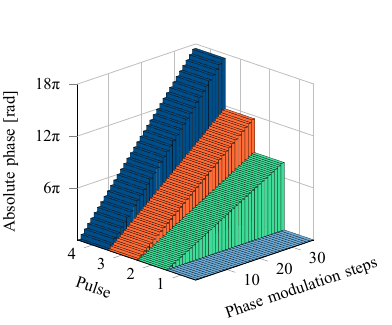}
\caption{Representation of phase modulation routine (left) and cogwheel phase cycling scheme (right) for N = 36 and n\textsubscript{1} = 0, n\textsubscript{2} = 4, n\textsubscript{3} = 6, and n\textsubscript{4} = 9 for a single pattern.}
\label{Fig2}
\end{figure}

By adopting this approach, it becomes feasible to prevent phase jumps that may lead to inaccurate values post-FT when repeatedly streaming the same pattern. This is because the phase shift for the i\textsuperscript{th} pulse between consecutive patterns will always remain equal to the respective $\Delta\Phi_\mathrm{i}$. This approach, known as phase modulation, has been recently addressed in terms of cogwheel phase cycling (COG) \cite{Jayachandran2024}. A comprehensive visual representation of this concept is illustrated in Fig. \ref{Fig2}. The values of N and n\textsubscript{i} are fundamental in defining the specific signals that can be retrieved, as well as the corresponding frequency at which each signal is detected. By carefully selecting these parameters, one can optimize the modulation scheme to enhance signal extraction. In our case, a particularly convenient set of values is N = 36, with integer divisors chosen as n\textsubscript{1} = 0, n\textsubscript{2} = 4, n\textsubscript{3} = 6, and n\textsubscript{4} = 9. This configuration ensures a well-defined phase evolution, facilitating precise frequency-domain analysis while maintaining coherence throughout the experimental sequence. The frequency associated to each linear and nonlinear signal will appear at the frequency given by:
\begin{equation}
\mathrm{f_i} = \frac{\mathrm{RepRate [Hz]}}{\mathrm{N}}\cdot\mathrm{n_i}
\label{Eq2}
\end{equation}

As highlighted in Table \ref{Tab1}, the chosen frequency combination allows to clearly separate all the linear components from the nonlinear ones. An alternative set of suitable parameters includes N = 40 with integer divisors selected as n\textsubscript{1} = 0, n\textsubscript{2} = 4, n\textsubscript{3} = 5, and n\textsubscript{4} = 8. However, conducting the experiment at a laser repetition rate of 4 kHz would result in the emergence of linear and nonlinear signatures at integer multiples of 100 Hz. This is suboptimal, since external frequencies from electrical instrumentation in the laboratory, usually operating at 50 Hz, could contaminate the signal with its harmonics, potentially compromise the accuracy of the measurements.

\begin{figure}[htbp]
\centering\includegraphics[]{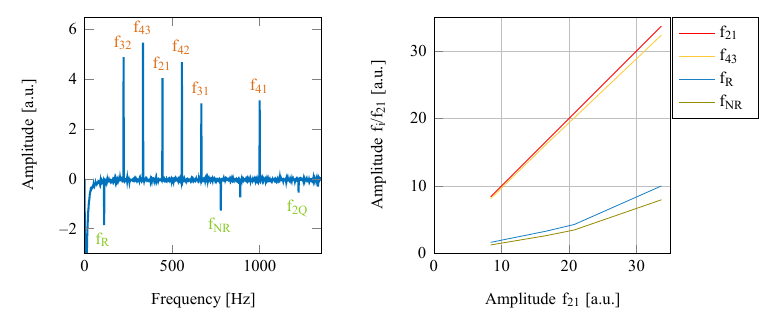}
\caption{On the left side, we assign the values from Table \ref{Tab1} to the peaks obtained from an experiment using the phase modulation routine performed with a laser repetition rate of 4 kHz, with N = 36 and n\textsubscript{1} = 0, n\textsubscript{2} = 4, n\textsubscript{3} = 6, and n\textsubscript{4} = 9. On the right side, it is reported the dependency of linear and nonlinear signals amplitude from the laser power for experiments performed using the same phase modulation scheme. The linear contributions f\textsubscript{21} and f\textsubscript{43} scale linearly with the laser power while the nonlinear features f\textsubscript{R} and f\textsubscript{NR} scale non-linearly.}
\label{Fig3}
\end{figure}

\begin{table}[htbp]
\caption{Phase modulation routine: individual phase shifts and signal frequencies$^{\alpha}$}
\label{Tab1}
\centering
\begin{tabular}{cccc}
\arrayrulecolor{black}\hline
f\textsubscript{i} & n\textsubscript{i} & $\Delta\Phi_\mathrm{i}$ & Frequency [Hz] \\
\arrayrulecolor{black}\hline
f\textsubscript{1} & 0 & $0\pi$   &\\
f\textsubscript{2} & 4 & $2/9\pi$ &\\
f\textsubscript{3} & 6 & $1/3\pi$ &\\
f\textsubscript{4} & 9 & $1/2\pi$ &\\
\arrayrulecolor{black}\hline
\rowcolor{option1!25}
f\textsubscript{21} & $4-0$ &  & 444.44 \\
\rowcolor{option1!35}
f\textsubscript{31} & $6-0$ &  & 666.66 \\
\rowcolor{option1!25}
f\textsubscript{41} & $9-0$ &  & 999.99 \\
\rowcolor{option1!35}
f\textsubscript{32} & $6-4$ &  & 222.22 \\
\rowcolor{option1!25}
f\textsubscript{42} & $9-4$ &  & 555.55 \\
\rowcolor{option1!35}
f\textsubscript{43} & $9-6$ &  & 333.33 \\
\arrayrulecolor{black}\hline
\rowcolor{mycolor4!25}
f\textsubscript{R}  & $-0+4+6-9$ &  & 111.11  \\
\rowcolor{mycolor4!35}
f\textsubscript{NR} & $-0+4-6+9$ &  & 777.77  \\
\rowcolor{mycolor4!25}
f\textsubscript{2Q} & $-0-4+6+9$ &  & 1222.22 \\
\arrayrulecolor{black}\hline
\end{tabular}

$^\alpha$Phase modulation routine performed with a laser repetition rate of 4 kHz, with N = 36 and n\textsubscript{1} = 0, n\textsubscript{2} = 4, n\textsubscript{3} = 6, and n\textsubscript{4} = 9. The first block delineates the modulation frequencies imparted to the phase of individual laser pulses. The orange and green blocks represent the linear and nonlinear contributions, respectively, specifying the frequencies at which the corresponding signals can be extracted.
\end{table}
Once these components are effectively separated in the frequency domain using the chosen parameters, their identity as linear or nonlinear can be unequivocally confirmed by their characteristic response to varying laser power. As illustrated in Fig. \ref{Fig3}, the amplitude of linear and nonlinear features in the FT peaks exhibits distinct scaling behaviours with respect to the laser power.

\begin{figure}[htbp]
\centering\includegraphics[]{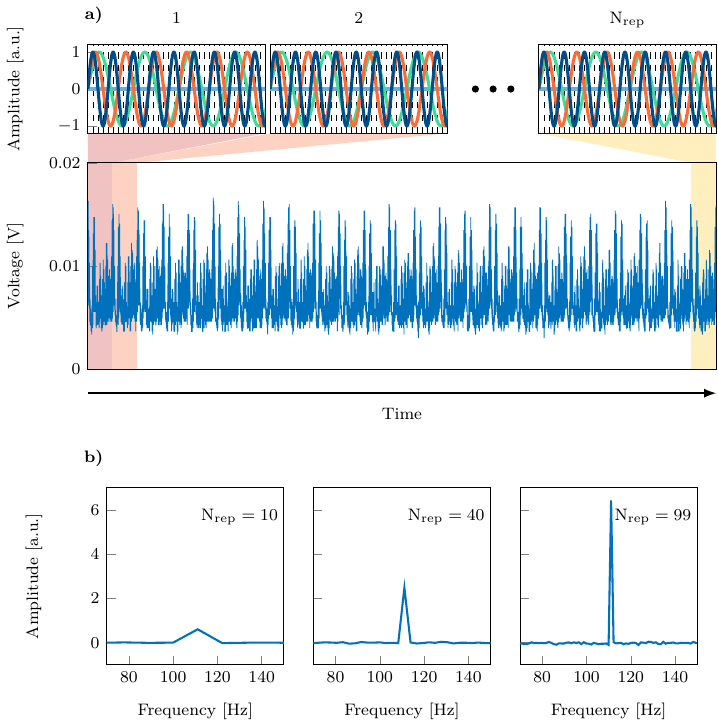}
\caption{Phase modulation schematic for N = 36 with parameters n\textsubscript{1} = 0, n\textsubscript{2} = 4, n\textsubscript{3} = 6, and n\textsubscript{4} = 9 for experiment recorded using a laser repetition rate of 4 kHz. \textbf{a)} Visualization of the pattern repetition during data acquisition. \textbf{b)} FT component for the rephasing peak at 111.11 Hz when the pattern N is repeated for N\textsubscript{rep} = 10, 40, and 99 times.}
\label{Fig4}
\end{figure}

Specifically, the linear contributions, such as f\textsubscript{21} and f\textsubscript{43}, follow a linear dependence on power, whereas the nonlinear features, denoted as f\textsubscript{R} and f\textsubscript{NR} scale non-linearly. Keeping in mind the limitations and problems that can arise from suboptimal combination of selected parameters N and n\textsubscript{i} with laser repetition rate and FIFO memory of the pulse shaper, the phase modulation (or phase cycling) can be employed to retrieve the nonlinear features at specific frequencies. Periodic streaming of the same pattern enhances the precision of the FT in resolving the frequency components contributing to the detected incoherent signal, as illustrated in Fig. \ref{Fig4}.

This approach effectively serves as an averaging procedure, facilitating the extraction of cleaner signal components during FT analysis. Indeed, if the modulation is performed for a short time that doesn’t allow sufficient repetitions of the pattern and therefore a poor data point collection, the FT will show broader peaks and lower signal to noise level.

\subsection{Sample device}
This study investigates a perovskite solar cell device employing a mixed-cation lead halide absorber layer with the composition (FAPbI\textsubscript{3})\textsubscript{0.992}(MAPbBr\textsubscript{3})\textsubscript{0.008}. The complete experimental methodology, encompassing material preparation, device architecture, and encapsulation protocols have been reported in a previously published work \cite{Zhang2024}. F-doped tin oxide (FTO) coated substrates were laser-etched or chemical etched with diluted HCl solution and cleaned by detergent, deionized (DI) water, acetone, and ethanol, which was further treated with UV/ozone (UVO) for 60 min prior to deposit SnO\textsubscript{2}. The SnO\textsubscript{2} film was prepared by chemical bath deposition (CBD) method. The CBD solution was prepared by mixing 1.25 g of urea (99.0\%, Sigma-Aldrich), 1.25 ml of HCl (37\%, Sigma-Aldrich), and 275 mg of SnCl\textsubscript{2}·2H\textsubscript{2}O (99.99\%, Sigma-Aldrich) in 100 ml of DI water TGA with stirring for 30 min.
The FTO glasses were dipped into the CBD solution and heated in oven at 90 \degree C for 4 hours. Then, the glasses were washed with DI and IPA for 15 min in sequence and annealed at 175 \degree C for 60 min. The FAPbI\textsubscript{3} based perovskite precursor solution was prepared by dissolving presynthesized FAPbI\textsubscript{3} (1.4 M), MACl (0.5 M), PbI\textsubscript{2} (0.13 M), and presynthesized MAPbBr\textsubscript{3} (0.012 M) in 1 ml of N,N-dimethylformamide (DMF, anhydrous, $\geq 99.8\%$, Sigma-Aldrich) and dimethyl sulfoxide (DMSO, anhydrous, $\geq 99.9\%$, Sigma-Aldrich) (8:1 v/v). The precursor solution was spin-coated on the SnO\textsubscript{2}-coated FTO substrate at 1000 rpm for 5 s and 5000 rpm for 20 s, where 1 mL of diethyl ether was dripped while spinning to form an adduct intermediate. The adduct films were annealed at 150 \degree C for 10 min and 100 \degree C for 10 min. The N2,N2,N2',N2',N7,N7,N7',N7'-octakis(4-methoxyphenyl)-9,9'-spirobi[9H-fluorene]-2,2',7,7'-tetramine (spiro-MeOTAD, 99\%, Merck) layer was formed on the perovskite layer by spin-coating of 30 \micro L of the solution comprising 60 mg spiro-MeOTAD in 0.7 mL chlorobenzene including 25.5 \micro L t-BP (4-tert-butylpyridine, 98\%, Sigma-Aldrich) and 15.5 \micro L Li-TFSI solution (540 mg Li-TFSI in 1 mL acetonitrile (99.8\%, Sigma-Aldrich)) at 4000 rpm for 20 s. All the processes are conducted under ambient condition. Finally, Au electrode with a thickness of about 70-100 nm was deposited on top of the spiro-MeOTAD layer by a thermal evaporator under $<2\times10^{-6}$ torr at an average rate of 0.5 \AA/s. Encapsulation was performed in a nitrogen-filled glovebox (O\textsubscript{2}/H\textsubscript{2}O levels $<0.1$ ppm) at room temperature using a two-part epoxy glue (Gorilla Epoxy). Prior to encapsulation, the devices were kept in the glovebox for a minimum of 30 minutes to ensure the removal of any residual solvents. A cover glass, uniformly coated with the epoxy, was placed on the entire active area, followed by the application of clamping pressure to eliminate trapped air. The encapsulated devices were subsequently stored in the glovebox for at least 12 hours to allow complete curing of the epoxy.

\section{Results and discussion}
\subsection{Trigger synchronization and filter out non-relevant data points}

\begin{figure}[htbp]
\centering\includegraphics[]{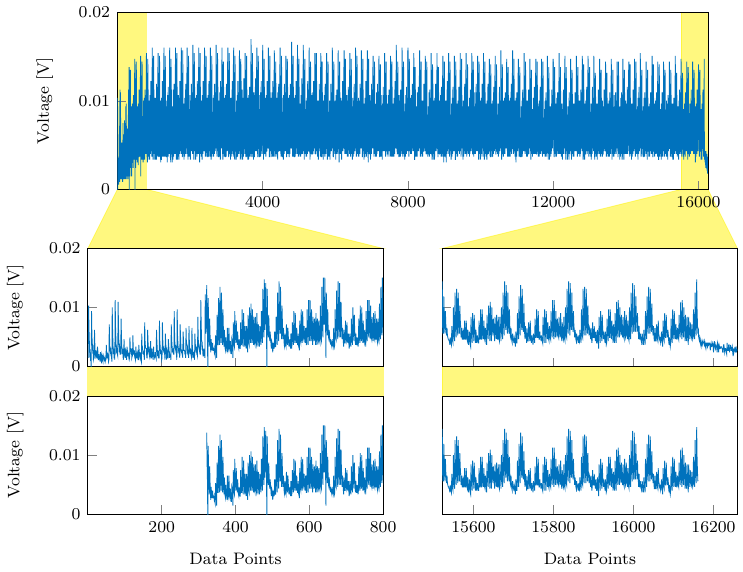}
\caption{Post-processing procedure. Given the exact number of collected data points, the initial points affected by pulse shaper streaming limitations are excluded, along with the final point marking the end of the phase modulation routine.}
\label{Fig5}
\end{figure}

The raw data, as initially recorded, lack the required cleanliness for direct FT processing. Two primary factors contribute to this: firstly, the need to eliminate additional data points collected during periods of Dazzler inactivity. Trigger synchronization has been introduced to significantly diminish the number of meaningless tail points at the end that are removed during the post-processing procedure; secondly, even highly performing electronic devices such as the Dazzler pulse shaper can suffer from electronical limitations. Since the laser repetition rate of 4 kHz is above the maximum supported streaming rate of the pulse shaper (1 kHz), the transmission speed of the waveforms from the computer to the radio-frequency generator (RF) is slower than the laser repetition rate. In few words, when a new pattern of waveforms is sent to the RF generator, as soon as the first waveform of the pattern is ready to be played in the RF generator, it is sent to the crystal at the next laser trigger event, but when the next laser trigger event occurs, it can be that the second waveform of the pattern is not ready to be played yet. Hence the RF generator has no other choice than to ignore the incoming trigger signals until the second waveform is ready. After that the new pattern has been fully loaded once, then, since the waveforms are stored in the memory, the repeated patterns are streamed without losing any trigger signals. Those missing triggers will result in an irregularity in the initial part of the collected signal which can affect the processed data. To avoid this inconsistency, the signal associated with the initial loading stage, which contains loading-related aberrations, has been removed during the post-processing procedure. Taking care of these details, it is possible to run a full experiment at higher repetition rate, speeding up the process while ensuring that the individual pulse energy is intense enough to work in the nonlinear regime.

Fig. \ref{Fig5} illustrates the necessary post-processing steps for optimal data analysis. The process of reshaping and trimming data during post-processing has a significant impact on the resulting 2D maps due to the high sensitivity of the FT to the number of data points included. Even the removal or addition of a few points can alter the FT outcome, potentially distorting spectral features and affecting subsequent data interpretation. One of the primary drawbacks of imprecise data collection and processing is the introduction of unintended phase shifts by the FT. These phase distortions can compromise the accuracy of the extracted spectral information, leading to misinterpretations of the underlying physical processes. Similar phase distortion-related effect has been already reported in other multidimensional spectroscopic techniques \cite{Khalil2003,Brixner2004,Anna2012}. To illustrate the importance of appropriate data processing, let’s consider the simulated photocurrent signal modulated by two sinusoidal frequencies displayed in Fig. 6 a), where the simulated data are recorded with a rate of 4 kHz, repeating a pattern N = 36 points for $\mathrm{N}_\mathrm{rep}>100$.

\begin{figure}[htbp]
\centering\includegraphics[]{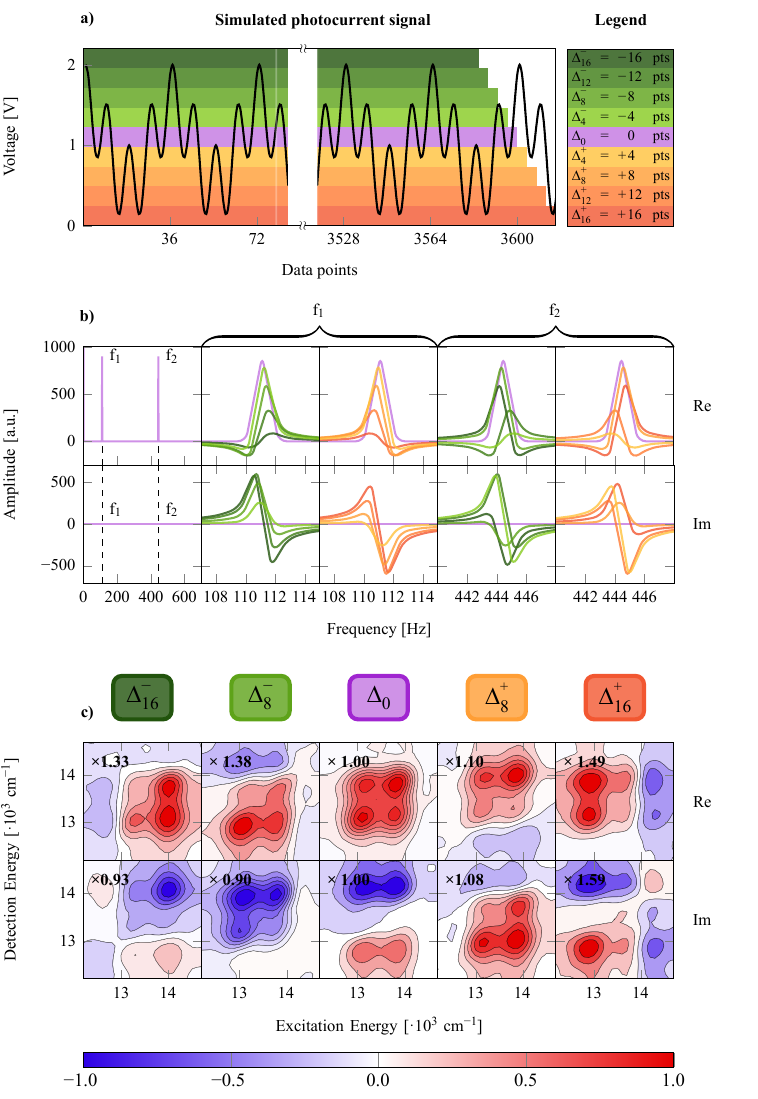}
\caption{Phase distortion and data handling. \textbf{a)} Simulated photocurrent signals. Each colour bar covers the amount of data points that are considered for FT: the purple bar refers to the optimal number of points to be considered ($\Delta_0$); the green bars refer to under-sampled data points ($\Delta_\mathrm{X}^-$) and the orange bars refer to over-sampled data points ($\Delta_\mathrm{X}^+$). The quantity X refers to the number of points that differ from the optimal case.}
\label{Fig6}
\end{figure}
\newpage
\begin{figure}[htbp]
\captionsetup{type=figure}
\caption*{\textbf{b)} FT signal showing the real (Re) and imaginary (Im) components at the two simulated frequencies (f\textsubscript{1} and f\textsubscript{1}) for the different trimming options addressed in panel a). The colour scheme corresponds to the legend in panel a). \textbf{c)} Real and imaginary 2D maps measured on solar cell device with subsequent trimming of the dataset at specific values given by $\Delta_{16}^-$, $\Delta_8^-$, $\Delta_0$, $\Delta_8^+$, and $\Delta_{16}^+$. The resulting picture illustrates the effect of data trimming on the spectral features. The 2D maps are visualized using eight equally spaced contour lines and each row is normalized in respect to $\Delta_0$.}
\end{figure}

\subsection{Selective and precise data-trimming for optimal FT operation}
Performing FT on the data simulated in time-domain allows to retrieve the information related to the contributing features in frequency-domain. Fig. \ref{Fig6} b) clearly visualizes the critical dependence of FT analysis on precise data acquisition and point selection. The FT inherently assumes periodic boundary conditions, meaning that an integer multiple of complete modulation cycles must be included in the dataset to maintain phase coherence. When this condition is met, the FT spectrum consists of well-defined, symmetric peaks centred at the characteristic modulation frequencies ($\Delta_0$). However, if the number of data points deviates even slightly—such as through the addition or removal of a few points—phase discontinuities arise at the signal boundaries. As can be seen from Fig. \ref{Fig6} c), the natural consequence of this phase leakage results in spectral distortions all-over the 2D maps, including peak broadening, asymmetric line shapes, and spurious frequency components. These distortions can compromise the integrity of spectral interpretation, particularly in high-precision nonlinear spectroscopic measurements such as A-2DES. To mitigate such effects, it is crucial to carefully control data acquisition parameters and ensure that the collected dataset aligns with integer multiples of the modulation period.

\subsection{Correction of spectral phase leakage resulting from population signal buildup}
\begin{figure}[htbp]
\centering\includegraphics[]{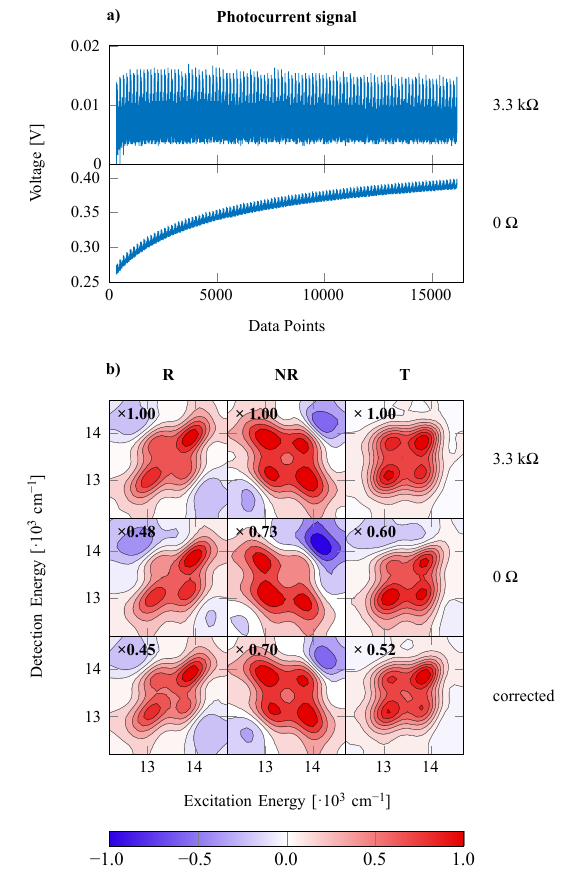}
\caption{Effect of population response accumulation on action-detected signal and its impact on the 2D spectra.\textbf{ a)} Measured photocurrent traces at two different circuit resistances: 3.3 k\ohm (upper trace) and 0 \ohm (lower trace). The absence of discharge results in signal accumulation over time. \textbf{b)} Corresponding 2D spectroscopic maps for rephasing (R), non-rephasing (NR), and absorptive (T) signals with 3.3 k\ohm resistance (1\textsuperscript{st} row), without discharge (2\textsuperscript{nd} row), and the discharged data after phase correction (3\textsuperscript{rd} row). The lack of discharge leads to significant phase distortion in the 2D spectra, which can be corrected through post-processing by applying phase factors $\exp(i\pi\phi_\mathrm{R})$ and $\exp(-i\pi\phi_\mathrm{R})$. The 2D maps are visualized using 8 equally spaced contour lines and each column is normalized in respect to the corresponding map in the first row.}
\label{Fig7}
\end{figure}

Appropriate phase correction algorithms may be required to minimize phase leakage and preserve the accuracy of the extracted spectral features \cite{Agathangelou2021,Sahu2023}. To illustrate this issue, we performed the same experiment on an identical sample under two different detection conditions. In one case, the signal was actively discharged between pulse sequences using a 3.3 k\ohm~resistance, while in the other case, no discharging tool was applied, leading to progressive signal accumulation. Fig. \ref{Fig7} a) shows the corresponding photocurrent traces recorded, that will correspond to a single pixel in the time-domain 2D map after performing FT. With a 3.3 k\ohm~resistance, the signal remains stable over time, while in the absence of discharge, the signal builds up. Fig. \ref{Fig7} b) presents the corresponding 2D maps under these two conditions in frequency-domain. When the signal is not discharged, strong phase distortions emerge, significantly affecting the spectral features. However, post-processing corrections can be applied to retrieve the correct information. By multiplying the rephasing and non-rephasing signals in the time-domain by phase factors $\exp(i\pi\phi_\mathrm{R})$ and $\exp(-i\pi\phi_\mathrm{NR})$, respectively, the phase distortions can be effectively removed, restoring the expected 2D spectral structure, as seen in the bottom row of Fig. \ref{Fig7} b). Importantly, the degree of deviation depends on the rate of signal discharge. The slower the discharge, the more pronounced the phase shift, as residual signals accumulate across multiple pulse sequences. This observation underscores the necessity of proper signal handling, particularly in high-repetition-rate experiments, where rapid population response decay is essential to ensure accurate phase-sensitive detection. The ability to correct such distortions post-experiment further highlights the robustness of phase-modulated detection techniques in preserving spectral integrity. For this reason, in this technique meticulous attention must be devoted to the generation of the four phase-modulated pulses, ensuring precise phase control throughout the experiment.

\subsection{Mitigate power-related nonlinear distortions introduced by the pulse shaper}
\begin{figure}[htbp]
\centering\includegraphics[]{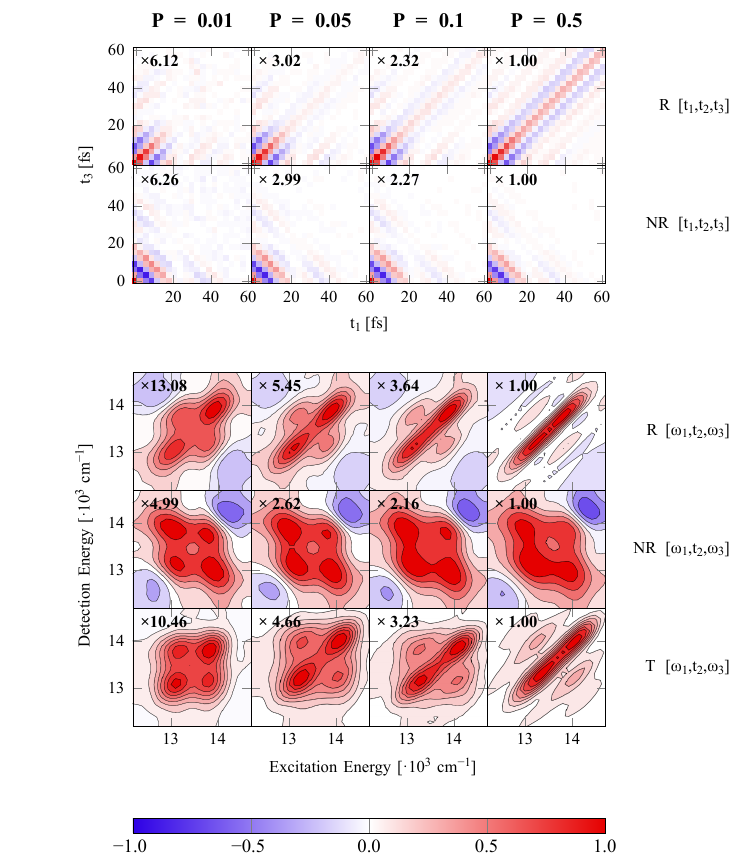}
\caption{2D data at population time $\mathrm{t}_\mathrm{2} = 100$ fs. Each column refers to an experiment operating under different streaming power of the Dazzler, highlighting distortion due to nonlinearities of the pulse shaper. The upper panel visualizes the rephasing (R) and non-rephasing (NR) maps in time-domain. The lower panel displays rephasing, non-rephasing and absorptive (T) maps in the frequency-domain. Each map is normalized to the corresponding map at P = 0.5 in the same row, displaying the scaling factor. The 2D maps are visualized using 8 equally spaced contour lines.}
\label{Fig8}
\end{figure}

The use of a Dazzler pulse shaper provides a reliable means of achieving phase-locked pulses through its intrinsic digital lock-in mechanism. However, to perform an experiment in the optimal conditions it is required careful management of multiple parameters to obtain reliable and reproducible data. One critical factor is the control over the power at which the Dazzler streams the acoustic waveform, as this can introduce aberrations into the 2D spectra due to nonlinearities in the RF generator, as illustrated in Fig.~\ref{Fig8} \cite{López-Ortiz2024}. Specifically, increasing the power of the acoustic waveform propagation within the crystal leads to the emergence of a distorted feature that manifests as an elongated background along the main diagonal of the rephasing spectra in time-domain. Therefore, this unwanted feature, arising from subtle imperfections in the pulse shaper, if uncorrected, can significantly distort signal intensities of the rephasing 2D map in frequency-domain, especially in off-diagonal peaks. As observed in our own dataset, the weights of these features can vary by more than a factor of two if the power intensity of the streamed waveform is not set properly. Beyond enhancing data quality, such details are essential to avoid potential misinterpretations, thus underscoring their value in pulse-shaper based spectroscopy. Notably, this anomalous contribution is observed exclusively in the rephasing contribution, whereas the non-rephasing component exhibits a nonlinear scaling with increasing intensity. The effect becomes already noticeable when the streaming power reaches approximately 4\% of the Dazzler’s maximum capability. Therefore, optimizing the experimental conditions requires a careful balance between maximizing nonlinear signal strength and avoiding anomalous features introduced by excessive power levels. Therefore, based on the specific properties of the sample, it is advisable to fine-tune the intensity of the Dazzler as high as possible while ensuring that the streaming power of the pulse shaper remains within the acceptable limits to prevent undesired distortions. This optimization is crucial for conducting experiments in a well-defined nonlinear regime while maintaining data integrity.

\section{Conclusions}
In this work, we have presented a detailed description and implementation of an A-2DES setup utilizing an acousto-optic pulse shaper. We focused on providing practical insights into the experimental methodology, particularly the implementation of phase modulation routines and the critical aspects of data acquisition and processing necessary for obtaining reliable and distortion-free 2D spectra. Our findings, demonstrated on perovskite solar cell device, highlight several crucial factors often overlooked in action-detected multidimensional spectroscopy. We showed that careful selection of phase modulation parameters (N, N\textsubscript{rep} and n\textsubscript{i} values) in conjunction with the laser repetition rate is essential to separate desired nonlinear signals from linear contributions and avoid interference from external noise sources. Furthermore, we identified and characterized significant limitations arising from FT processing due to improper data handling, residual population effects in high-repetition-rate experiments, and nonlinearities introduced by the pulse shaper itself at higher streaming powers. We have demonstrated effective post-processing strategies to mitigate these shortcomings, such as precise data trimming and phase correction procedure, ensuring the integrity of the extracted spectral information. The distinct power-scaling behaviour observed for linear and nonlinear signal components further underscores the need for careful optimization of pulse shaper and laser parameters to operate effectively within the nonlinear regime while minimizing aberrations. This comprehensive methodological overview and distortion analysis provides a valuable guide for researchers employing pulse-shaper-based A-2DES. The ability to reliably acquire and process A-2DES data paves the way for deeper investigations into the complex ultrafast charge carrier and energy transfer dynamics within advanced materials and operational optoelectronic devices.

\begin{backmatter}
\bmsection{Acknowledgment}
We would like to thank Dr. T. Pascher for helping with the detection system. E.A. and T.P. acknowledge financial support from the Swedish Energy Agency grant 50709-1, VR grant 2021-05207, Olle Engkvist foundation grant 235-0422, the European Union’s Horizon 2020 research and innovation program under the Marie Skłodowska-Curie grant agreement no. 945378 and no. 871124 Laserlab-Europe, the SNC Fellowship Program in Korea 2023, and the Royal Physiographic Society of Lund. L.B. and N.F.v.H. acknowledge support through the MCIN/AEI Projects PID2021-123814OB-I00, TED2021-129241BI00, CEX2019-000910-S, Fundacio Privada Cellex, Fundacio Privada Mir-Puig, and the Generalitat de Catalunya through the CERCA program. N.F.v.H. acknowledges support from ERC Advanced Grant 101054846-FastTrack. S.-H.L. and N.-G.P. acknowledge financial support from the National Research Foundation of Korea (NRF) through grants funded by the Korean government (MSIT) under contract number NRF-2021R1A3B1076723 (Research Leader Program).

\bmsection{Disclosures}
The authors declare no conflicts of interest. The manuscript reflects only the authors view, the European Union and the Research Executive Agency are not responsible for any use that may be made of the information it contains.

\bmsection{Data availability}
Data underlying the results presented in this paper are not publicly available at this time but may be obtained from the authors upon reasonable request.
\end{backmatter}
\bibliography{Main}






\end{document}